# Double Dirac Point Semimetal in Two-Dimensional Material: Ta$_2$Se$_3$

Yandong Ma[†,⊥*], Yu Jing[†,⊥], and Thomas Heine[†,⊥*]

[†] Wilhelm-Ostwald-Institut für Physikalische und Theoretische Chemie, Universität Leipzig, Linnéstr. 2, 04103 Leipzig, Germany

[⊥] Department of Physics and Earth Sciences, Jacobs University Bremen, Campus Ring 1, 28759 Bremen, Germany

*Corresponding author: myd1987@gmail.com (Y.M.); thomas.heine@uni-leipzig.de (T.H.)

Here, we report by first-principles calculations one new stable 2D Dirac material, Ta$_2$Se$_3$ monolayer. For this system, stable layered bulk phase exists, and exfoliation should be possible. Ta$_2$Se$_3$ monolayer is demonstrated to support two Dirac points close to the Fermi level, achieving the exotic 2D double Dirac semimetal. And like 2D single Dirac and 2D node-line semimetals, spin-orbit coupling could introduce an insulating state in this new class of 2D Dirac semimetals. Moreover, the Dirac feature in this system is layer-dependent and a metal-to-insulator transition is identified in Ta$_2$Se$_3$ when reducing the layer-thickness from bilayer to monolayer. These findings are of fundamental interests and of great importance for nanoscale device applications.

TOC Figure

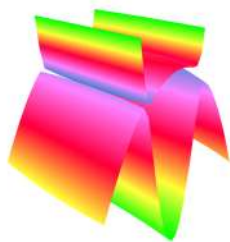

Double Dirac Points

# I. Introduction

Since the discovery of graphene [1], two-dimensional (2D) Dirac materials have attracted tremendous interest owing to their intriguing electronic properties as well as being desirable to combine their exotic behaviors with nanoelectronic devices. Currently, a variety of 2D materials have been proposed to harbor the single Dirac semimetal phase. These include graphene [1], silicene [2,3], germanene [2,3], stanene [4], graphene allotropes [5-7], *Pmmn*-boron [8], janugraphene and chlorographene [9]. In the absence of spin-orbit coupling (SOC), the combination of inversion and time-reversal symmetries protects the single Dirac point in graphene [10], as in similar materials. SOC introduces a gap in the single Dirac cones, giving rise to a topological insulating phase [11,12]. The discovery of topological quantum phase promoted the rise of node-line semimetals, which host Dirac line nodes when SOC is neglected [13-15]. 2D node-line (NL) semimetal was shown to exist in $Hg_3As_2$ monolayer with the HK lattice [16]. Like 2D single Dirac point (SDP) semimetals, SOC opens a band gap at the Dirac node line, leading to a 2D topological crystalline insulating state [16]; see **Figure 1**. For these two classes of 2D Dirac materials, strong SOC in heavy-element compounds is crucial for the related exotic features.

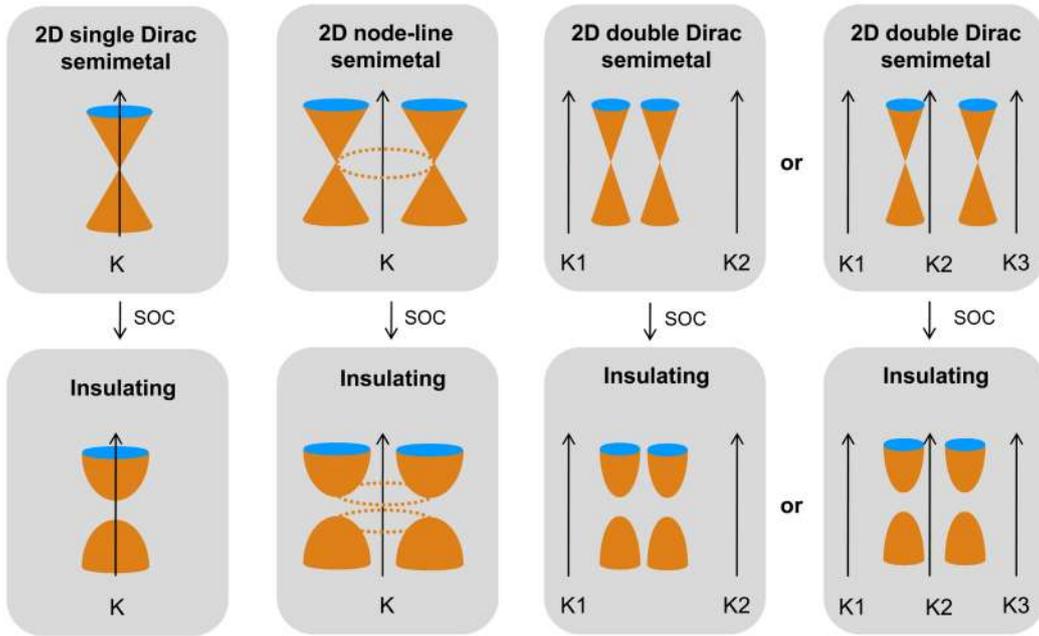

**Figure 1**. Schematic diagrams of 2D single Dirac point semimetal, 2D node-line semimetal and 2D double Dirac point semimetal, and their evaluations under SOC. Here, the K and Kn (n=1,2,3) denote the high symmetry *k* points in the first Brillouin zone. Edge states are not shown for clarity.

Besides SDP and NL semimetals, the concept of double Dirac point (DDP) semimetal has also been very recently introduced [17]. Distinct from SDP, which only has one single Dirac point, and NL semimetal, whose band crossing points around the Fermi level forms a closed loop (Dirac line node), DDP semimetals host two Dirac points which are located at different positions, as shown in **Figure 1**. This new phase of quantum matter presents a significant expansion of Dirac materials beyond SDP and NL semimetals, and new opportunities to explore exotic physics [17]. However, compared with the extensive research in 2D SDP and 2D NL semimetals, few 2D, or even three-dimensional (3D), materials have been shown to be in the DDP semimetal class [17]. Very recently, the theoretically proposed structure, 2D ionic boron, was shown to have double Dirac points [18]; but the Dirac points locate above the Fermi level, and synthesis is expected to be challenging and has not been reported to date. No real 2D DDP semimetal has been predicted at all, which makes studying their potential exotic physics difficult. The search for 2D DDP semimetals, specifically in the naturally existed systems, thus is of fundamental scientific interest and technological significance.

Here, using first-principles calculations, we explore the electronic properties of one so far unnoticed 2D material, $Ta_2Se_3$ monolayer. We demonstrate that $Ta_2Se_3$ monolayer exhibits double Dirac points within the first Brillouin zone, achieving the intriguing 2D DDP semimetal state without any additional tuning. We further show that, like 2D SDP and 2D NL semimetals, SOC introduces a gap at the Dirac cones near the Fermi level, leading to an insulating phase with a band gap of 48 meV. Furthermore, we show that the Dirac features in $Ta_2Se_3$ are layer-dependent and an intriguing layer-dependent metal-to-insulator transition is obtained in $Ta_2Se_3$ when thinning down to the monolayer. $Ta_2Se_3$ monolayer can likely be manufactured (75 meV/Å$^2$) by micromechanical cleavage or liquid exfoliation in experiments since its bulk phase is layered material in nature [19]. This makes $Ta_2Se_3$ monolayer a most promising candidate material for investigating 2D double Dirac physics.

**II. Results and Discussion**

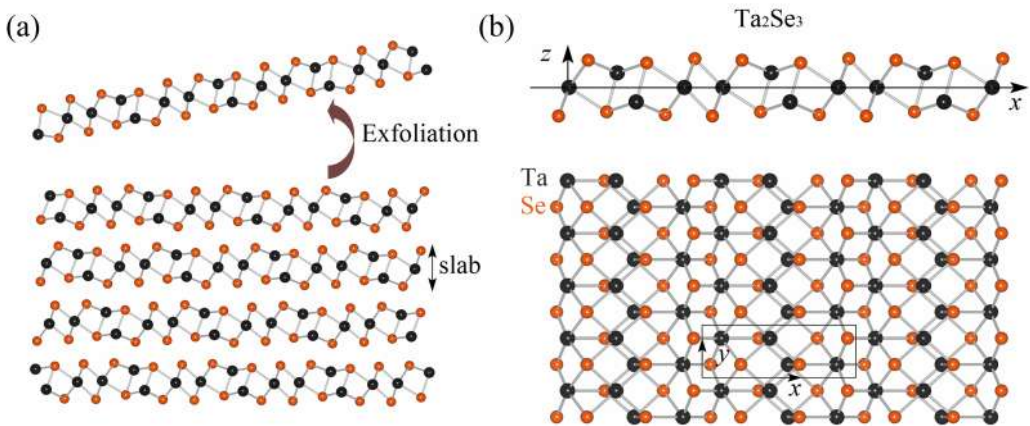

**Figure 2**. (a) Schematic illustration of $Ta_2Se_3$ monolayer isolation from the bulk surface, simplified by a five-layer slab. (b) Side (upper panel) and top (low panel) views of the crystal structures of Ta$_2$Se$_3$ monolayer. The solid lines denote the unit cells.

Our first-principles calculations are performed within density functional theory (DFT) using the VASP code [20,21]. For details, please see the Supporting Information. Bulk Ta$_2$Se$_3$ has been known since 1968, and has been synthesized by heating the mixed elements to 1140 °C and quenching to room temperature [19]. Bulk Ta$_2$Se$_3$ crystallizes in the *P21/m* space group, and it is a layered compound with two slabs in one unit cell. In each slab, the Ta atoms roughly form a layer, which is sandwiched between two Se layers. Due to the strong chemical bonding within each slab, the adjacent slabs are weakly coupled. Recent experimental studies have well established that layered materials can be made into stable monolayer structures using micromechanical cleavage or liquid exfoliation [22,23]. Specifically, liquid exfoliation is even suitable for layered compounds with strong interlayer interactions [24]. We have estimated the cleavage energy for Ta$_2$Se$_3$ monolayer following the approach as illustrated in **Figure 2a**, and obtain 75 meV/Å$^2$, a relatively large, but not untypical value for two-dimensional crystals (the smallest cleavage energy estimations on grounds of comparable level of theory are 14 and 26 meV/Å$^2$ for graphite and MoS$_2$, respectively [25]). Therefore, we expect that the experimental fabrication of Ta$_2$Se$_3$ monolayer is possible, in particular given the recent progress in liquid phase exfoliation [26]. **Figure 2(b)** shows the crystal structure of Ta$_2$Se$_3$ monolayer. It displays a tetragonal lattice and contains two formula units in each unit cell. Inversion symmetry holds well for this system. The optimized lattice constants are a=9.855 Å, b=3.378 Å. For more details about the monolayer structure, please see the Supporting Information.

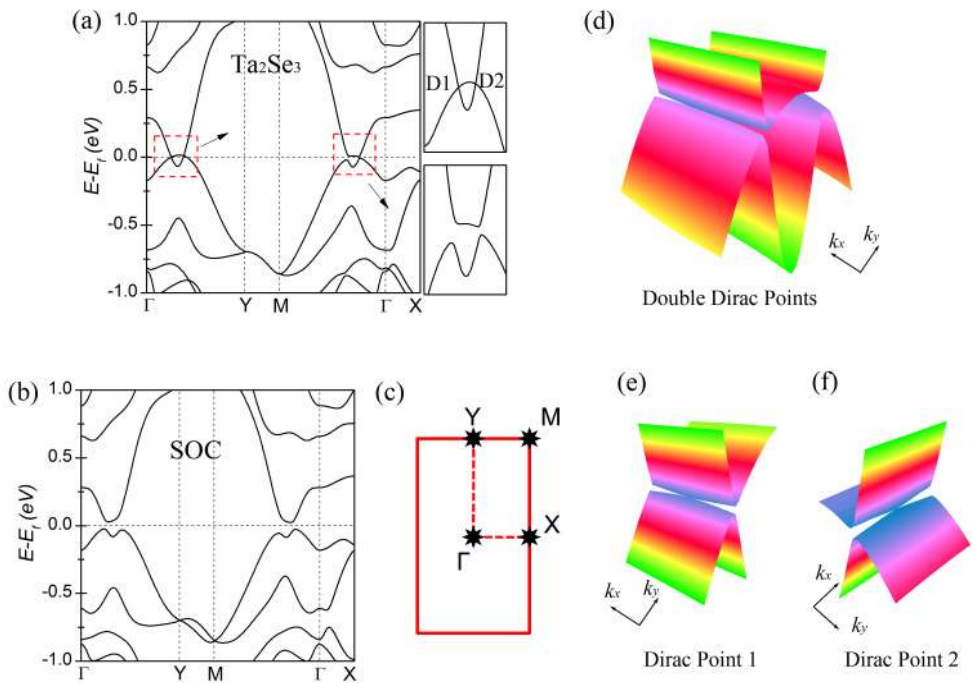

**Figure 3**. Electronic structure of Ta$_2$Se$_3$ monolayer (a) without and (b) with SOC. (c) 2D Brillouin

zone of Ta$_2$Se$_3$ monolayer. (d-f) Band dispersions of Ta$_2$Se$_3$ monolayer around Fermi level in 2D $k$ space ($k_x$ and $k_y$) with energy as the third dimension.

The band structure of Ta$_2$Se$_3$ monolayer without SOC is shown in **Figure 3(a)**. The bands near the Fermi level are located around the Γ point, which quickly disperse by more than 0.5 eV towards the Brillouin zone edge. The corresponding band structure magnifications near the Fermi level are plotted in the right panels of **Figure 3(a)**. Clearly, there is a band inversion along the Γ-Y direction, giving rise to two band crossing points locating close to the Fermi level. Unlike the case along the Γ-Y direction, the bands along the Γ-M direction display semiconducting behavior with an indirect band gap of 48 meV. This indicates that Ta$_2$Se$_3$ monolayer behaves like a semimetal with double band crossing points. It is worthy emphasizing that the similar two band crossing points feature already exists in 2D NL semimetals [16], however, in that case the node lines have different momenta directions, while in the present 2D DDP case they have the same direction in k space. To understand the electronic properties of Ta$_2$Se$_3$ monolayer, we show the band dispersions of Ta$_2$Se$_3$ monolayer around the crossing points in 2D $k$ space in **Figure 3(d)-(f)**. Excitingly, the band crossing points in Ta$_2$Se$_3$ monolayer are Dirac points. What is special here is that Ta$_2$Se$_3$ monolayer hosts double Dirac points. This is in sharp contrast to the case of 2D NL semimetal in which the band crossing between the conduction band and the valence band around the Fermi level forms a closed nodal ring. This is also significantly different from the 2D SDP semimetals [1-3] which only have one single Dirac point (**Figure 1**). The appearance of double Dirac points suggests that Ta$_2$Se$_3$ monolayer belongs to another class of 2D Dirac materials, namely the rarely obtained 2D DDP semimetals.

Commonly, 2D Dirac materials could be driven into nontrivial insulating phases by including SOC. For example, 2D SDP semimetals display a Dirac point in the absence of SOC, but exhibiting a topological insulating behavior when including SOC [11,12]. Moreover, 2D NL semimetal possesses Dirac line nodes when SOC is neglected, and SOC opens a gap at the Dirac node line, leading to a 2D topological crystalline insulating state [13-15]. The common feature between them is that SOC plays an important role in gaping the band crossing between the valence and conduction bands and thus driving the quantum phase transition from semimetal to insulator. Concerning the strong SOC strength within Ta and Se atoms, it is natural to wonder how SOC affects the Dirac state of Ta$_2$Se$_3$ monolayer. The band structure of Ta$_2$Se$_3$ monolayer with SOC is shown in **Figure 3(b)**. Noteworthy, a situation very similar to that in 2D SDP and 2D NL semimetals is observed in Ta$_2$Se$_3$ monolayer: after taking SOC into account, the valence and conduction bands of Ta$_2$Se$_3$ monolayer move away from each other and a gap is opened; see **Figure 3(b)**. The SOC-induced band gap is about 48 meV. This is identified as a SOC-induced phase transition from 2D DDP semimetal to 2D insulator. Given the fact that opening a gap by SOC in 2D SDP and 2D NL semimetals can result in a nontrivial insulating phase—topological insulator or topological crystalline insulator, we also expect to obtain nontrivial insulating phase in 2D DDP semimetal. To this end, we analyze the band topology of Ta$_2$Se$_3$ monolayer and find that the insulating state is neither topological insulator nor topological crystalline insulator. The concept of DDP semimetal was very recently introduced and the underlying physics is still under exploration [17]. However, we believe that Ta$_2$Se$_3$ monolayer may belong to one unknown nontrivial insulating phase. And the material realization of DDP semimetal in Ta$_2$Se$_3$ monolayer provides the first step for theoretical and experimental investigations of such new state. To get deeper insight into the properties of Ta$_2$Se$_3$ monolayer, we investigate its edge states by constructing a nanoribbon with symmetric edges. The width of the nanoribbon is selected to be 7.7 nm to avoid spurious interactions between the two edges. The edge band structure is shown in **Figure S1**. Interestingly, we find that there are two nearly-flat edge states. And the edge states display weakly dispersing feature near the $\overline{M}$ point. The existence of such exotic edge states probably indicates a nontrivial state in Ta$_2$Se$_3$ monolayer. And future works on this topic are needed.

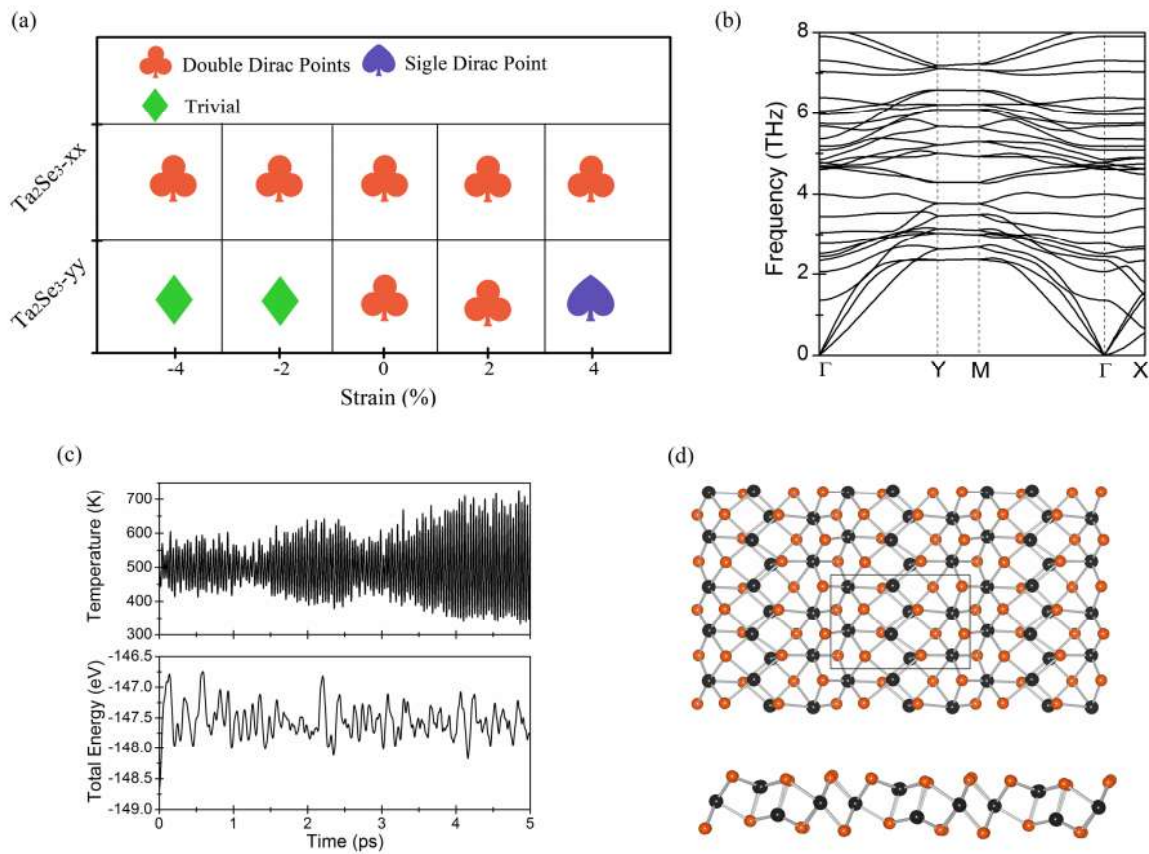

**Figure 4**. (a) Electronic characterization of Ta$_2$Se$_3$ monolayer under influence of strain. In the suffixes of the labels, 'xx' and 'yy' denote that the strain is applied along the x and y axes, respectively. (b) Phonon dispersions of $Ta_2Se_3$ monolayer. (c) Fluctuations of temperature and total energy with time obtained from MD simulation of $Ta_2Se_3$ monolayer at 500 K. (d) Snapshot of $Ta_2Se_3$ monolayer at the end of MD simulation at 500 K from side and top views. The black and orange balls in (d) denote Ta and Se atoms, respectively.

As 2D materials, the electronic structure of $Ta_2Se_3$ monolayer is expected to be tunable by applying strain. Concerning the structural anisotropy, we impose separately uniaxial strains along x- and y-axes on the 2D plane, and the results are summarized in **Figure 4(a)**. We emphasize that the discussion on compressive strain here is more about academic interest rather than for practical applications because of its difficulties in experimental realization. As seen in **Figure 4(a)**, the compressive and tensile strains along x-axis can reach 4% without deforming the double Dirac points of $Ta_2Se_3$ monolayer. And the detailed features are similar to the unstrained case, i.e., the two Dirac points are close to Fermi level and hence $Ta_2Se_3$ monolayer is still a 2D DDP semimetal, and importantly the global band gap can be obtained with including SOC, giving rise to an insulating phase [see **Figure S2** and **S3**]. Things are different along the y-axis. When 4% tensile strain along y-axis is applied, the double Dirac points of $Ta_2Se_3$ monolayer is destroyed, and a new single Dirac point is formed along the Γ-Y direction, see **Figure S4**. On the other hand, when compressive strain along y-axis is larger than 2%, the double Dirac points of $Ta_2Se_3$ monolayer is also vanished and $Ta_2Se_3$ monolayer is a metal and a semiconductor under 2% and 4% cases respectively. Therefore, as shown in **Figure 4(a)**, by applying strain along y-axis, the fascinating phase transition from trivial to DDP state and then to SDP state can be obtained in $Ta_2Se_3$ monolayer. And except for the case of 4% tensile strain, under the strain along y-axis, $Ta_2Se_3$ monolayer is always semiconducting due to SOC; see **Figure S5**. Generally speaking, application of external strain can effectively modify the electronic properties of $Ta_2Se_3$ monolayer.

Next, we study the electronic properties of $Ta_2Se_3$ with increasing layer thickness. To this end, we consider the bilayer, trilayer, four-layer, five-layer and bulk structures, which are obtained from the corresponding bulk phase. As seen in **Figure S6**, when stacking $Ta_2Se_3$ monolayers together to form a multilayer film, the double Dirac feature is destroyed, which is mainly due to the interlayer interaction. Moreover, as shown in **Figure S6-S8**, the multilayer and bulk structures of $Ta_2Se_3$ are metals regardless of SOC. Therefore, by increasing layer number from monolayer to multilayer or bulk, a transition from DDP semimetal (insulating) to metallic states with excluding (including) SOC can be established in $Ta_2Se_3$. It is important to note that compared with the layer-thickness-dependent indirect-to-direct-gap transition observed in $MoS_2$ and $WS_2$ [27,28], the layer-thickness-dependent metal-to-insulator transition obtained here are expected to be more exciting. It will be very interesting to observe such transition in experiments. These structures thus can be used as a platform to investigate different kinds of electronic states controlled by strain and layer-thickness.

In what follows, we assess the stability of $Ta_2Se_3$ monolayer. The dynamic stability of $Ta_2Se_3$ monolayer is first investigated by calculating its phonon dispersions, and the corresponding results are presented in **Figure 4(b)**. The absence of imaginary phonon modes in the whole Brillouin zone confirms that $Ta_2Se_3$ monolayer is dynamically stable. To further examine its thermal stability, we perform the MD simulations for this monolayer. As seen in **Figure 4(d)**, no structure reconstruction can be observed after heating at 500 K for 5 ps with a time step of 1 fs, firmly suggesting that $Ta_2Se_3$ monolayer is thermally stable. The corresponding changes of energy and temperature with time during the simulations are also shown in **Figure 4(c)**. The stability analysis makes us more confident to believe that $Ta_2Se_3$ can form stable monolayer structure, and further experimental studies are strongly called for. We also extend our investigation to the $Ta_2Te_3$ monolayer and find that the unique DDP feature is absent in $Ta_2Te_3$ monolayer as it supports a single Dirac point below the Fermi level; for details, please see the corresponding discussion in Supporting Information.

## Ⅲ. Conclusion

In summary, we demonstrate a new stable 2D material, $Ta_2Se_3$ monolayer, which we expect that it can be fabricated from its bulk phase in experiments. It exhibits exciting electronic properties: $Ta_2Se_3$ monolayer is a 2D DDP semimetal. Particularly, the identified notable 2D DDP semimetal state represents a significant advance in the field of 2D Dirac materials, and moreover, SOC can introduce an insulating state. We also demonstrate that by applying strain, the Dirac states in this material can be effectively controlled. In addition, we reveal that the Dirac features in both systems are layer-dependent. And more interestingly, the layer-dependent metal-to-insulator transition is identified in $Ta_2Se_3$ when reducing the thickness down to monolayer, which arises from the quantum confinement effect.

## Supporting Information

The computational methods, electronic structure of $Ta_2Se_3$ nanoribbon, electronic structures of $Ta_2Se_3$ monolayer without and with SOC under various strains, electronic structures of multilayer and bulk $Ta_2Se_3$ without and with SOC, the corresponding results and discussions of $Ta_2Te_3$ monolayer, phonon dispersions and MD results of $Ta_2Te_3$ monolayer, and the geometric structures of $Ta_2Se_3$ and $Ta_2Te_3$ monolayers. This material is available free of charge via the Internet.

## Acknowledgement

Financial support by the Deutsche Forschungsgemeinschaft (DFG, HE 3543/18-1, HE 3543/27-1)

are gratefully acknowledged.

**Additional information**

Competing financial interests: The authors declare no competing financial interests.